\documentclass[lnbip]{svmultln}
\usepackage[numbers]{natbib}
\usepackage{xcolor}
\usepackage{booktabs}
\usepackage{tabularx}
\usepackage{array}
\usepackage{multirow}
\usepackage{enumitem}
\usepackage{seqsplit}
\usepackage{graphicx}
\usepackage{subcaption}
\usepackage{makeidx}  
\usepackage[subtle]{savetrees}
\begin{document}
\mainmatter              
\title{IOTEL: A Tool for Generating \\ IoT-enriched Object-Centric Event Logs}
\titlerunning{A Tool for Generating IoT-enriched Object-Centric Event Log}  
%
\author{Jia Wei \and Xin Su \and Chun Ouyang}
\authorrunning{Jia Wei et al.}   
%
%
\institute{Queensland University of Technology, Brisbane, Australia}

\maketitle              

\begin{abstract}        
Integrating Internet of Things (IoT) data with business process event logs is crucial for analysing IoT-enhanced processes, yet remains challenging due to differences in abstraction levels and the separation of data sources. Simply incorporating raw IoT data increases the size and complexity of the resulting log, often requiring additional processing before process analysis can be performed. While tools for generating IoT-enriched event logs exist, they either rely on specialised schemas or focus on extracting event logs from sensor data, offering limited support for integrating process-relevant IoT data into existing event logs.
To address this gap, we present IOTEL, a tool for systematically generating IoT-enriched object-centric event logs (OCEL). By building on the OCEL schema, IOTEL enables structured IoT data integration compatible with existing process mining tools. It support practitioners and researchers in analysing IoT-enhanced business processes, as demonstrated in a real-world scenario. A video demonstrating the tool is available online.\footnote{\url{https://github.com/XinSu869/IOTEL/tree/main}}
\end{abstract}
\vspace{-2.5\baselineskip}
\section{Introduction}
\label{sec:intro}
\vspace{-0.5\baselineskip}
The Internet of Things (IoT) has been widely adopted in various industries. By embedding sensors into physical assets and processes, organisations can obtain fine-grained operational data and real-time insights~\citep{bendaya2021}. However, understanding how IoT technologies influence business process execution and performance remains challenging.
This is mainly because analysing IoT-enhanced processes requires integrating IoT data with process execution data, the latter typically recorded in event logs. IoT data consists of low-level sensor measurements originating from physical systems, whereas event logs capture high-level records of process execution maintained in enterprise systems~\citep{ValderasTS22}. This mismatch in abstraction levels, combined with the separation of data sources, makes direct integration complex. Naively incorporating raw IoT data into event logs unnecessarily increases the size and complexity of the resulting log, necessitating substantial preprocessing before they can support process analysis. Effective integration, therefore, requires identifying and incorporating only process-relevant IoT data.

Despite the recognised importance of integrating IoT data with event log, dedicated tools for generating IoT-enriched event logs remain limited. The BROOM toolbox~\citep{imenkamp2025broom} supports IoT-enhanced process mining, producing data structured in accordance with the CORE metamodel~\citep{bertrand2025object}. Consequently, generating IoT-enriched event logs with BROOM requires prior familiarity with this specialised schema. As a result, tools that enable the systematic integration of process-relevant IoT data into event logs without prescribing a specific schema for the resulting logs remain scarce.

To address this gap, we present IOTEL, a tool for integrating process-relevant IoT data into object-centric event logs (OCEL). OCEL captures processes involving multiple interacting business objects, which aligns naturally with IoT-enhanced processes where devices interact with multiple objects during execution. By building directly on the OCEL schema, IOTEL enables IoT data enrichment while preserving the underlying log structure, thereby facilitating compatibility with existing process mining tools. With IOTEL, business users and process analysts can integrate business data and IoT data without developing custom data integration pipelines. It also provides the research community with a practical means of generating IoT-enriched event logs to support analysis of IoT-enhanced business processes. We demonstrate the feasibility of IOTEL through a real-world scenario.

\vspace{-1\baselineskip}

\section{Background and Related Work}
\label{sec:bg}
\vspace{-0.5\baselineskip}
This section first introduces the OCEL log schema and relevant IoT ontologies for specifying the semantic structure of IoT data, and then reviews tools for generating IoT-enriched event logs.
\vspace{-0.7\baselineskip}
\subsection{Object-centric Event Log}
\vspace{-0.5\baselineskip}
Object-centric event logs (OCEL 2.0)~\citep{berti2024ocel} have been proposed to represent real-world business processes that involve multiple object types interacting in complex ways over time. For instance, in a cargo pickup process, multiple object types such as trucks, cargo, silos, and pickup plans interact with each other. A pickup plan may involve multiple trucks, while a truck can participate in different pickup plans over time. OCEL enables capturing these complex object interactions together with their associated process activities in a unified log.
OCEL 2.0 supports several storage formats. In this work, we adopt the relational database representation, where objects and events are stored in dedicated tables and linked through event-to-object and object-to-object relations tables. This relational structure provides the foundation for integrating IoT data into the OCEL log.

\vspace{-0.7\baselineskip}
\subsection{IoT Data Ontologies}
\vspace{-0.5\baselineskip}
The rapid expansion of IoT deployments has led to an increasing volume of heterogeneous data, which in turn has driven the development of ontologies that provide semantic models of IoT data. Among these, the W3C’s Semantic Sensor Network Ontology (SSN)\footnote{\url{https://www.w3.org/TR/vocab-ssn/}} and its lightweight core, Sensor, Observation, Sample, and Actuator Ontology (SOSA)\footnote{\url{https://www.w3.org/TR/vocab-ssn-2023/}}, are widely adopted for describing data generated through sensing and actuation processes.
SOSA organises descriptions around key elements such as sensors, observations, samples, features of interest, and actuators. Its lightweight structure supports interoperability across diverse IoT domains and enables consistent semantic annotation. SSN builds upon SOSA by providing a more detailed conceptual model that supports richer descriptions required in scientific and engineering applications.
Several extensions further adapt these models to specific focuses. IoT-Lite~\citep{bermudez2016iot} offers a simplified subset of SSN concepts tailored to lightweight deployments and data analytics purposes. Domain-specific ontologies such as MASON and FTOnto~\citep{gruger2023iot} extend SSN/SOSA concepts to represent IoT data generated in manufacturing and production processes.

\vspace{-0.7\baselineskip}
\subsection{Existing Tools for IoT-Enriched Event Log Generation}
\vspace{-0.5\baselineskip}
Only a limited number of tools have been proposed to generate IoT-enriched event logs. Brzychczy et al.~\citep{brzychczy2025iot} introduce a tool for generating event logs from industrial sensor data. Similarly, Seiger et al.~\citep{SEIGER2026101870} propose a method to abstract high-level process activities from IoT sensor streams. However, both approaches focus on deriving process event data directly from low-level sensor data.
In contrast, our work considers IoT data as external information that can be integrated with existing business process event logs to enhance process analysis.
A highly related tool is the BROOM toolbox\footnote{\url{https://github.com/chimenkamp/IOT-PM-Suite/tree/main}}, which provides support for IoT-enhanced process mining. While it includes a module for generating IoT-enriched event logs, it prescribes the CORE metamodel~\citep{bertrand2025object} as the underlying log schema, requiring prior knowledge of this meatamodel before the tool can be effectively applied.
\vspace{-1\baselineskip}
\section{Tool Description}
\label{sec:tool}
\vspace{-0.5\baselineskip}

\begin{figure}[ht!!]
    \centering
    \vspace{-1\baselineskip}
    \includegraphics[width=\linewidth]{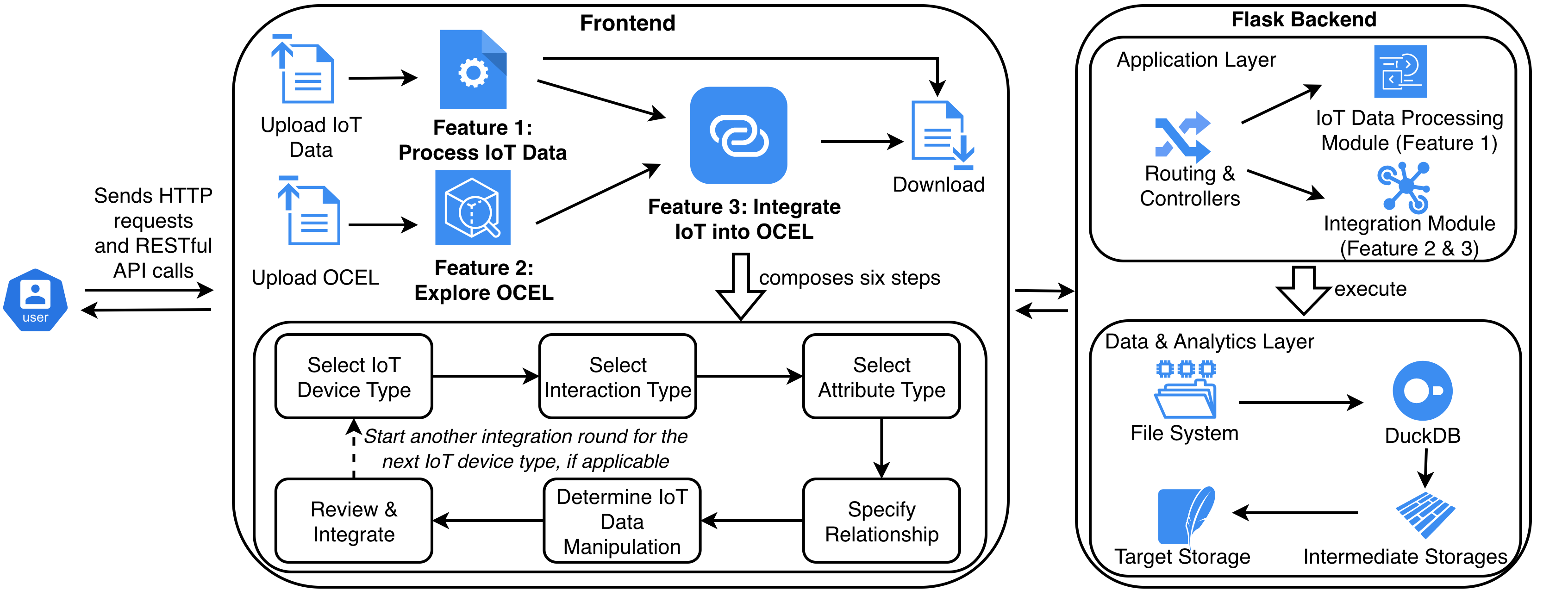}
    \caption{Architecture of IOTEL}
    \vspace{-2\baselineskip}
    \label{fig:IOTEL_Architecture}
\end{figure}

In this work, we develop IOTEL, a semi-automated tool for integrating process-relevant IoT data into object-centric event logs, based on the method proposed in (\citep{quteprints262989}, Chapter~5).
Fig.~\ref{fig:IOTEL_Architecture} presents the overall architecture of the tool. IOTEL provides three core features: IoT data processing, OCEL log exploration, and IoT data integration. In the following sections, we first introduce these features from the frontend perspective, followed by a description of the backend architecture.

\vspace{-1\baselineskip}
\subsection{IoT Data Processing}
\vspace{-0.5\baselineskip}
Since the primary objective of IOTEL is to enable the systematic integration of IoT data into OCEL logs, the first feature is IoT data processing. Raw IoT data is typically characterised by high volume, velocity, and heterogeneity~\citep{gruger2023iot}. Although IoT ontologies such as SOSA/SSN provide rich semantic models for representing IoT data, directly incorporating all IoT data into OCEL logs would introduce unnecessary semantic complexity and substantially increase log size.
Rather than applying post-integration filtering, IOTEL adopts a pre-integration processing strategy in which only process-relevant IoT data is retained prior to OCEL log enrichment. To achieve this, we derive a subset ontology derived from SOSA/SSN ontologies, specifically tailored to the integration of process-relevant IoT data. The subset is extracted through exclusion and inclusion criteria, summarised in Table~\ref{tab:ontology_scope}. Based on these criteria, the following ontology elements are retained to capture process-relevant IoT data: \texttt{sosa:Sensor}, \texttt{sosa:Actuator}, \texttt{sosa:FeatureOfInterest}, \texttt{sosa:ObservableProperty}, \texttt{\seqsplit{sosa:ActuatableProperty}}, \texttt{ssn:Property}, \texttt{sosa:Result}, \texttt{sosa:Platform}, \texttt{ssn:Deployment}, and \texttt{\seqsplit{sosa:resultTime}}.

\begin{table}[h!!!]
\vspace{-0.8\baselineskip}
\caption{Criteria for Selecting SOSA/SSN Ontology Elements for Process-Relevant IoT Data Representation}
\label{tab:ontology_scope}
\centering
\vspace{-1\baselineskip}
\small
\setlength{\tabcolsep}{4.5pt}
\renewcommand{\arraystretch}{0.95}
\begin{tabularx}{\linewidth}{p{1.3cm}|p{3.2cm}|X}
\hline
\textbf{Criteria} & \textbf{Decision} & \textbf{Rationale} \\
\hline
\multirow{2}{*}{\shortstack{Exclusion \\ Criteria}}
& Remove relationship predicates (e.g., \texttt{sosa:hasResult})
& These relationships are not required at the IoT data layer, as they are re-established during the integration phase with OCEL objects and events. \\
\cline{2-3}

& Remove procedure-related classes (e.g., \texttt{sosa:Observation}, \texttt{sosa:Procedure})
& These describe sensing execution rather than IoT data values to be recorded in the event log. \\
\hline

\multirow{7}{*}{\shortstack{Inclusion \\ Criteria}}
& Include classes representing objects monitored or affected by IoT devices
& Integration first requires identifying the business object that the IoT device monitors or affects during the process execution. \\
\cline{2-3}

& Include classes or properties describing observed attributes and produced values
& Each IoT interaction must specify the affected object attribute and yield a recorded value, whether single or aggregated. \\
\cline{2-3}

& Include classes capturing spatial information
& Location information captures where the interactions between IoT devices and business processes occur, i.e., a single event, multiple events, or a full process span. \\
\cline{2-3}

& Include classes capturing temporal information
& Temporal information captures when the interactions between IoT devices and business processes occur. \\
\cline{2-3}

& Include classes capturing IoT devices
& Identifying the generating device ensures traceability and prevents ambiguity when multiple sensors operate concurrently. \\
\cline{2-3}

& Include classes capturing IoT infrastructure
& Information about platforms and device capabilities can be valuable when assessing how IoT infrastructure contributes to process execution, for example, when devices operate interdependently with others. \\
\hline

\end{tabularx}
\vspace{-1\baselineskip}
\end{table}

\begin{figure}[htbp]
    \centering
    \vspace{-1.8\baselineskip}
    \begin{subfigure}{0.49\textwidth}
        \centering
        \includegraphics[width=\linewidth]{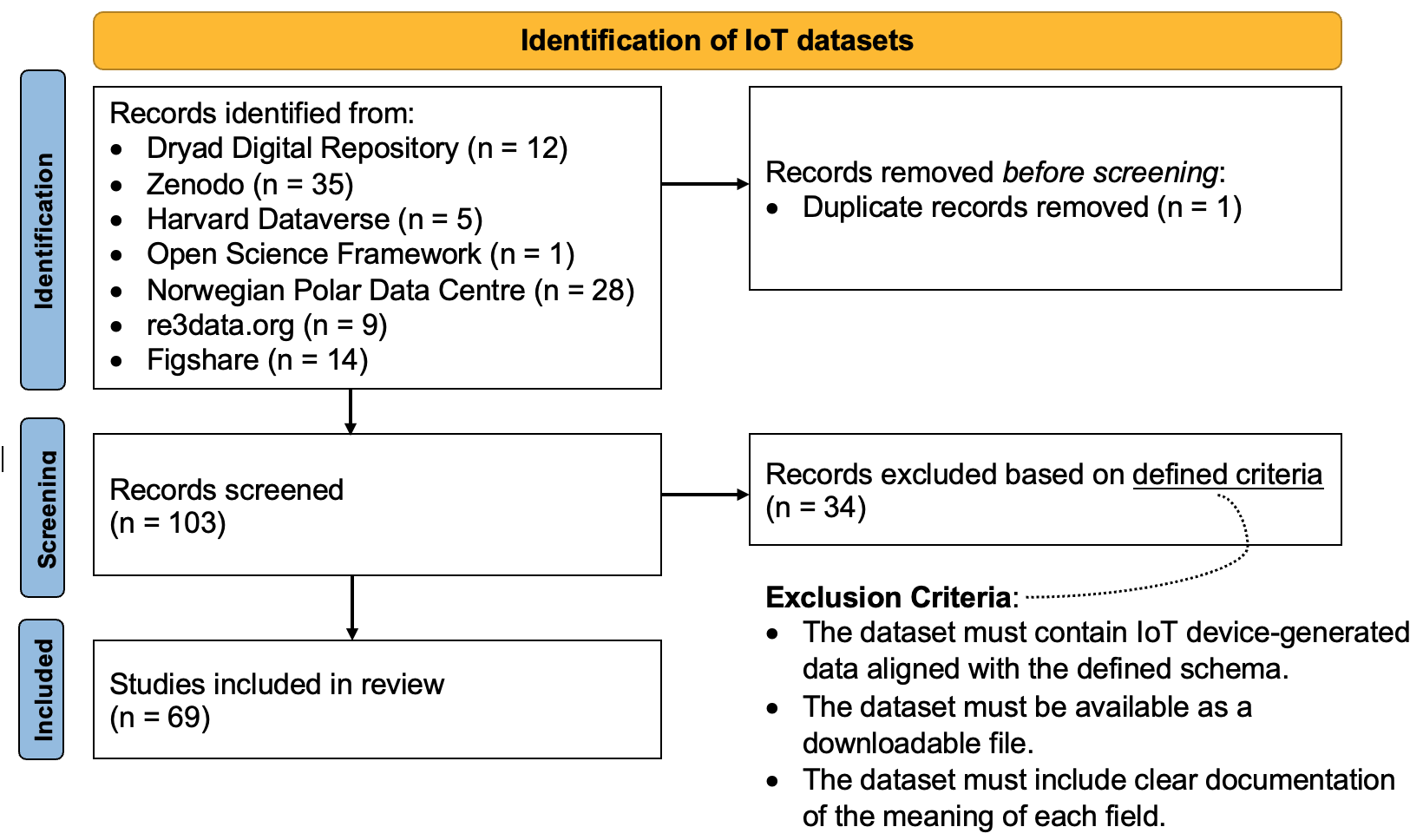}
        \caption{\scriptsize{The Process of Identifying IoT Datasets}}
        \label{fig:prisma}
    \end{subfigure}
    \hfill
    \begin{subfigure}{0.49\textwidth}
        \centering
        \includegraphics[width=\linewidth]{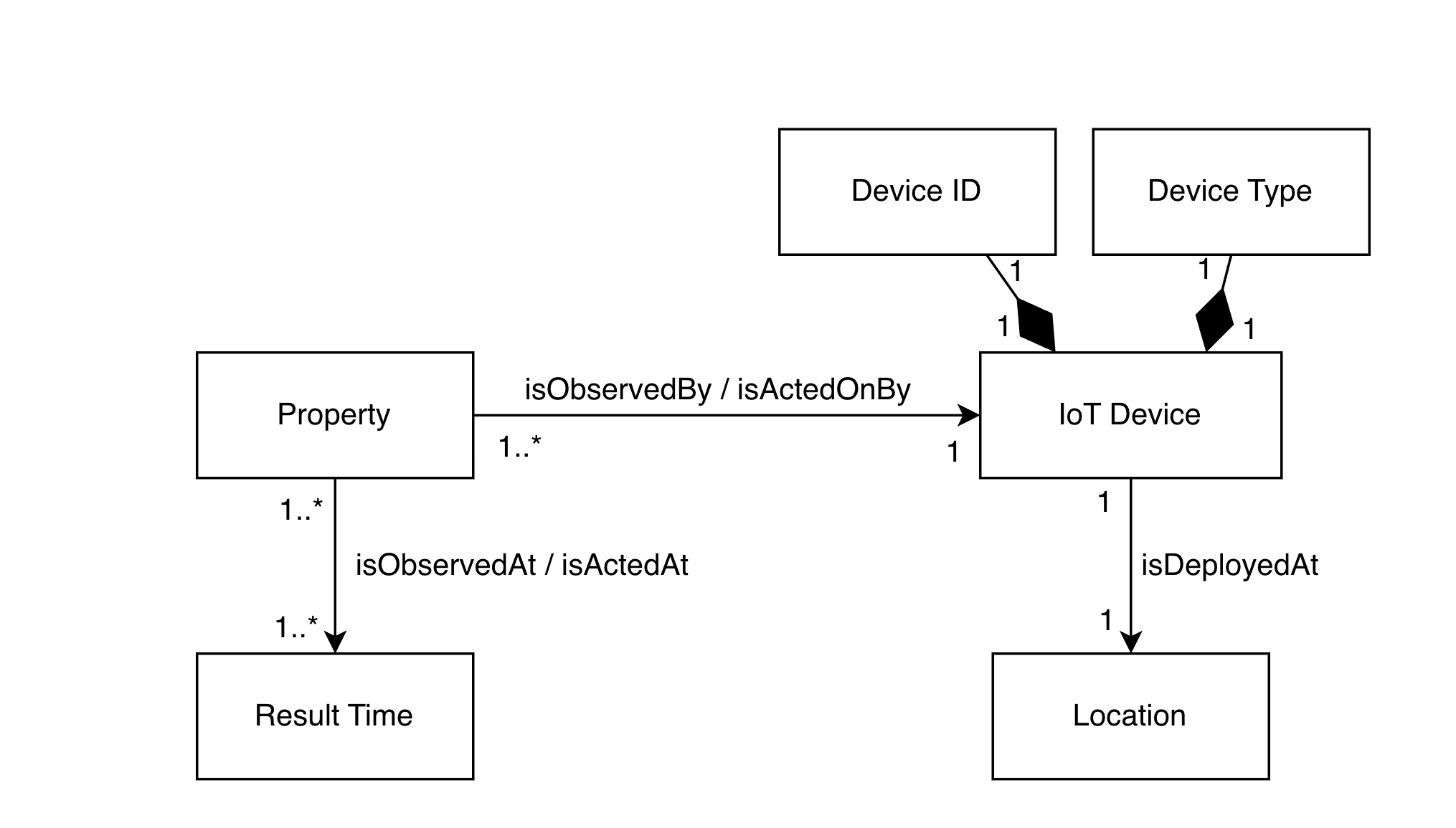}
        \caption{\scriptsize{Derived Subset of SOSA ontology for IoT data Processing}}
        \label{fig:schema}
    \end{subfigure}
    \vspace{-0.5\baselineskip}
    \caption{Overview of Dataset Selection and Derived Schema for IoT Data Processing toward OCEL Integration}
    \label{fig:combined}
    \vspace{-2\baselineskip}
\end{figure}

To evaluate the applicability of this subset, we analysed publicly available IoT datasets and mapped their data structures to the derived schema. In total, 69 datasets satisfying the selection criteria were identified (Fig.~\ref{fig:prisma}). Detailed information about these datasets is available online\footnote{\url{https://github.com/XinSu869/IOTEL/tree/main/common_iot_schema_verification}}. Each dataset was instantiated against the derived schema to verify that its core semantics could be consistently represented.
The final schema used for IoT data pre-processing is shown in Figure~\ref{fig:schema}. Sensor and actuator concepts are unified under a common \texttt{Device} abstraction, where each device is associated with a unique identifier and a device type (e.g., sensor or actuator). Observable and actuatable properties are consolidated into a unified \texttt{Property}, depending on the device type. Since both Platform and Deployment capture location-related semantics, they are abstracted into a common \texttt{Location} concept. The FeatureOfInterest class is excluded, as it represents the business object with which the IoT device interacts, and this information is already captured within the OCEL event log.

\vspace{-0.5\baselineskip}
\subsection{OCEL Log Exploration}
\vspace{-0.7\baselineskip}

The second feature focuses on exploring the OCEL log, which captures the execution of the business process. Specifically, we conduct statistical analysis to present key characteristics of the log, including the number of events, objects, activities, object types, and their relationships. This step is critical, as the OCEL log serves as the structural foundation for subsequent IoT data enrichment.

\vspace{-0.5\baselineskip}
\subsection{OCEL Integration with IoT Data}
\vspace{-0.7\baselineskip}
The final feature integrates the processed IoT data into the OCEL log by determining how the data should be represented, either as object attributes or as event attributes. This decision is guided by the interaction pattern between IoT devices and process activities or business objects.

IoT devices may participate in process execution in different ways. When a device is used at a specific activity without being associated with a business object, its data are recorded as event attributes. When a device is used in activities involving a business object, the representation depends on whether the measured property describes an attribute of the object. For instance, a weighing sensor capturing both empty and loaded truck weights reflects properties of the truck itself, and its data are stored as object attributes. When IoT devices continuously generate data that may affect the processes of multiple object types, such as weather conditions affecting truck and cargo processes, domain knowledge determines whether the data should be attached to specific activities or represented as object attributes.

Beyond attribute-level enrichment, the user specifies the qualifier defining the relationship between IoT devices and process elements. The tool also supports configurable IoT data manipulation options, including aggregation functions (e.g., minimum, maximum, average, and median), filtering based on user-defined value ranges, or retaining the raw data without processing. After reviewing the configuration, the IoT-enriched OCEL log is generated and made available for download. The procedure can be repeated for additional IoT device types.

\vspace{-0.5\baselineskip}
\subsection{Backend Architecture}
\vspace{-0.7\baselineskip}

The backend comprises two principal modules: the IoT data processing module and the integration module, which are routed through dedicated controller files. The IoT data processing module identifies process-relevant IoT data, while the integration module supports OCEL log exploration and integrates the processed IoT data into the log. The execution of these modules is underpinned by four foundational engines, detailed as follows:

\vspace{-0.7\baselineskip}
\subsubsection{File System}
A structured local file system acts as the primary artefact repository, systematically categorising resources into distinct directories. This includes `uploads/' for raw input files, `adjusted/' for intermediate data mapping outputs, and `processed/' for the merged, analytical datasets.

\vspace{-0.7\baselineskip}
\subsubsection{DuckDB}
Acting as the core analytical processing engine, DuckDB executes highly optimised SQL queries directly against intermediate data files. This embedded approach ensures robust, stateless data transformations whilst mitigating the extensive memory overhead typically associated with loading massive datasets into memory.

\vspace{-0.7\baselineskip}
\subsubsection{Intermediate Storages (Parquet Files)}
To guarantee computational efficiency, the system utilises the Parquet file format for intermediate data representation. Its columnar storage paradigm is inherently optimised for the complex analytical queries and aggregations continually demanded by the DuckDB engine during data processing.

\vspace{-0.7\baselineskip}
\subsubsection{Target Storage (OCEL SQLite Database)}

An SQLite database, strictly adhering to the OCEL 2.0 schema, functions as the ultimate data repository. The overarching architecture supports non-destructive, additive schema migrations, ensuring that newly integrated IoT attributes and auxiliary relationship tables are seamlessly aggregated without compromising the provenance of pre-existing events.

\vspace{-1\baselineskip}
\section{Maturity}
\label{sec:maturity}
\vspace{-0.5\baselineskip}

IOTEL is currently available as an implemented prototype and has been evaluated in a real-world use case. This section first presents the prototype’s graphical user interface, followed by a description of the application scenario.

\vspace{-0.5\baselineskip}
\subsection{Prototype}
\begin{figure}[h!]
    \centering
    \vspace{-1.7\baselineskip}
    \includegraphics[width=\linewidth]{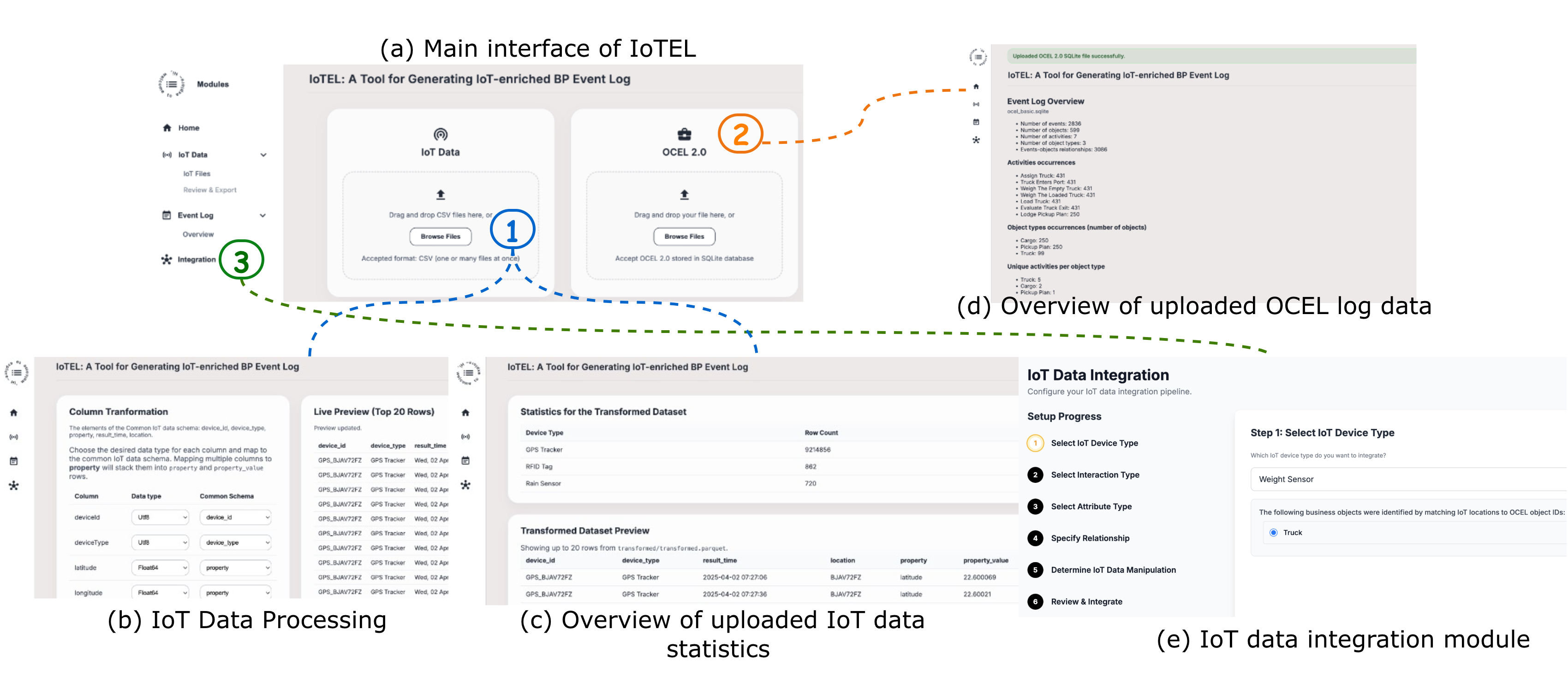}
    \vspace{-1.8\baselineskip}
    \caption{A graphical view of annotated IOTEL GUIs}
    \label{fig:GUI}
    \vspace{-2.5\baselineskip}
\end{figure}

\vspace{-0.7\baselineskip}
The proposed features have been implemented in an interactive prototype with a graphical view of annotated IOTEL GUIs, as shown in Fig.~\ref{fig:GUI}. 
The tool provides dedicated interfaces corresponding to the three core features described earlier.
Fig.~\ref{fig:GUI}(a) presents the main interface, structuring the tool around IoT data processing, OCEL log exploration, and IoT data integration. Fig.~\ref{fig:GUI}(b) and (c) correspond to the IoT data processing feature, supporting schema-based preprocessing and providing statistical summaries of the processed dataset, including detected device types and their record counts. Fig.~\ref{fig:GUI}(d) shows the OCEL log exploration interface, offering descriptive statistics on process activities and business objects, as well as a process map derived from the OCEL data. Fig.~\ref{fig:GUI}(e) presents the integration interface, enabling step-by-step configuration of the integration process, including the selection of IoT device types, interaction patterns, attribute mappings, and relationships between IoT data and OCEL log elements.
Overall, the prototype demonstrates that the proposed features are fully implemented and executable in an interactive environment. Implementation details, usage instructions, and a short video demonstrating the prototype are available in the project's GitHub repository\footnote{\url{https://github.com/XinSu869/IOTEL/tree/main}}.

\vspace{-0.5\baselineskip}
\subsection{Application}
\vspace{-0.5\baselineskip}

To demonstrate practical feasibility, the IOTEL prototype was applied in a real-world industrial scenario involving IoT-enhanced cargo pickup operations at a port. The port faces challenges in detecting fraudulent activities, particularly those involving abnormal truck weight manipulation~\citep{app10031056}. To address this issue, IoT technologies such as GPS tracking and smart weight sensors were deployed to monitor truck movements and weight deviations in real time.
However, process data and IoT data were stored separately. Operational data (e.g., truck entry times, gate information, and weighing records) resided in the port management system, while IoT data were maintained in dedicated IoT databases. Integrating these heterogeneous data sources was necessary to analyse how IoT information enhances process transparency and performance.
Using IOTEL, raw IoT data (e.g., GPS and weight sensor records) were preprocessed and integrated with the OCEL event log through the configurable mapping interface, resulting in an IoT-enriched OCEL log for analysis. Due to data confidentiality, synthetic datasets simulating the real-world scenario are also provided in the GitHub repository. The application confirms the practical feasibility of integrating IoT data with OCEL event logs using the prototype.

\section{Conclusion}
\label{sec:conclusion}
\vspace{-0.5\baselineskip}
In this work, we present IOTEL, a semi-automated tool for generating IoT-enriched event logs to support the analysis of IoT-enhanced business processes. IOTEL is designed to assist both practitioners and researchers by enabling the structured integration of process-relevant IoT data into OCEL log.
The current implementation assumes that an OCEL log is available as input. Transforming enterprise system data into the OCEL format is therefore outside the scope of this work and can be achieved using existing tools that support OCEL log extractions. Furthermore, IOTEL currently focuses on tabular IoT data, reflecting common representations aligned with SOSA/SSN concepts. Future work will extend IOTEL to support additional data modalities, such as sensor stream and multimodal IoT data, and investigate systematic approaches for integrating such data into event logs.


\vspace{-1.5\baselineskip}

%
%
\bibliographystyle{splncs04}
\bibliography{citations}

@article{bendaya2021,
author = {Mohamed Ben-Daya, Elkafi Hassini, Zied Bahroun and Bayan H. Banimfreg},
title = {The role of internet of things in food supply chain quality management: A review},
journal = {Quality Management Journal},
volume = {28},
number = {1},
pages = {17--40},
year = {2021},
publisher = {Taylor \& Francis}
}

@article{ValderasTS22,
  author       = {Pedro Valderas and
                  Victoria Torres and
                  Estefan{\'{\i}}a Serral},
  title        = {Modelling and executing {IoT}-enhanced business processes through {BPMN}
                  and microservices},
  journal      = {J. Syst. Softw.},
  volume       = {184},
  pages        = {111139},
  year         = {2022}
}

@phdthesis{quteprints262989,
           title = {Integrating IoT Data into Business Process Event Logs for Context-aware Analytics},
          author = {Jia Wei},
           month = {},
          school = {Queensland University of Technology},
            year = {}
}

@article{berti2024ocel,
  title={{OCEL (Object-Centric Event Log) 2.0 Specification}},
  author={Berti, Alessandro and Koren, Istv{\'a}n and Adams, Jan Niklas and Park, Gyunam and Knopp, Benedikt and Graves, Nina and Rafiei, Majid and Li{\ss}, Lukas and others},
  journal={arXiv preprint arXiv:2403.01975},
  year={2024}
}

@article{gruger2023iot,
  title={{IoT-enriched} event log generation and quality analytics: a case study},
  author={Gr{\"u}ger, Joscha and Malburg, Lukas and Bergmann, Ralph},
  journal={it-Information Technology},
  volume={65},
  number={3},
  pages={128--138},
  year={2023},
  publisher={De Gruyter Oldenbourg}
}

@inproceedings{bermudez2016iot,
  title={{IoT-Lite}: a lightweight semantic model for the Internet of Things},
  author={Bermudez-Edo, Maria and Elsaleh, Tarek and Barnaghi, Payam and Taylor, Kerry},
  booktitle={2016 INTL IEEE conferences on uic/atc/scalcom/cbdcom/iop/smartworld},
  pages={90--97},
  year={2016},
  organization={IEEE}
}

@article{brzychczy2025iot,
  title={{IoT} Miner: Intelligent Extraction of Event Logs from Sensor Data for Process Mining},
  author={Brzychczy, Edyta and Jessen, Urszula and Kluza, Krzysztof and Sriram, Sridhar and Nettelnstroth, Manuel Vargas},
  journal={arXiv preprint arXiv:2509.05769},
  year={2025}
}

@article{SEIGER2026101870,
title = {A domain-specific language and architecture for detecting process activities from sensor streams in IoT},
journal = {Internet of Things},
volume = {36},
pages = {101870},
year = {2026},
issn = {2542-6605},
author = {Ronny Seiger and Daniel Locher and Marco Kaufmann and Aaron F. Kurz}
}

@article{bertrand2025object,
  title={An object-centric core metamodel for {IoT}-enhanced event logs},
  author={Bertrand, Yannis and Imenkamp, Christian and Malburg, Lukas and Ehrendorfer, Matthias and Franceschetti, Marco and Gr{\"u}ger, Joscha and Leotta, Francesco and Mangler, J{\"u}rgen and Seiger, Ronny and Koschmider, Agnes and others},
  journal={arXiv preprint arXiv:2506.21300},
  year={2025}
}

@article{imenkamp2025broom,
  title={BROOM: Toolbox for {IoT-Enhanced} Process Mining},
  author={Imenkamp, Christian and Bertrand, Yannis and Gr{\"u}ger, Joscha and Malburg, Lukas and Franceschetti, Marco and Ehrendorfer, Matthias and Koschmider, Agnes},
  year={2025}
}

@Article{app10031056,
AUTHOR = {Song, Rongjia and Huang, Lei and Cui, Weiping and Óskarsdóttir, María and Vanthienen, Jan},
TITLE = {Fraud Detection of Bulk Cargo Theft in Port Using Bayesian Network Models},
JOURNAL = {Applied Sciences},
VOLUME = {10},
YEAR = {2020},
NUMBER = {3},
ARTICLE-NUMBER = {1056},
ISSN = {2076-3417}
}







%
\end{document}